\documentstyle[12pt]{article}

 \def\Th{\vartheta}
\def\g{\gamma} 
\def\a{\alpha}
\def\b{\beta}

\def\k{\kappa}
\def\l{\lambda} 
\def\m{\mu}
\def\n{\nu}
\def\f{\phi} 
\def\x{\chi}
\def\p{\pi} 
\def\r{\rho}
\def\s{\sigma}

\def\o{\omega} 
\def\be{\begin{equation}}
\def\bea{\begin{eqnarray}}
\def\nn{\nonumber}
\def\ee{\end{equation}}
\def\eea{\end{eqnarray}}
\def\wt{\widetilde}
\def\abs#1{|#1|}
\def\half{{\textstyle{1\over 2}}}
\def\sign{\mathop{\rm sign}}

\begin{document}

\title{Collisions of Einstein-Conformal Scalar Waves}
\author{{\em C.~Klim\v c\'\i k}~\thanks{Theory Division of the Nuclear Centre,
Charles University, V Hole\v sovi\v ck\'ach 2,
Prague, Czechoslovakia}~~\thanks{e-mail: presov @
cspuni12.bitnet}
\and{\em P.~Koln\'\i k}~$^*$~\thanks{e-mail: kolnik @ cspuni12.bitnet}}
\date{}
\maketitle
\begin{abstract}
A large class of solutions of the Einstein-conformal scalar equations in D=2+1
and D=3+1 is identified. They describe the collisions of asymptotic
conformal scalar waves and are generated from Einstein-minimally coupled
scalar spacetimes via a (generalized) Bekenstein transformation. Particular
emphasis is given to the study of the global properties and the singularity
structure of the obtained solutions. It is shown, that in the case of the
absence of pure gravitational radiation in the initial data, the formation
of the final singularity is not only generic, but is even inevitable.
\vskip 2cm
\rightline{PRA-HEP-92/19~~~~~~~~}
\vskip 1cm
\end{abstract}

\newpage

\renewcommand\Large{\large}

\section{Introduction}

Colliding gravitational waves have attracted a lot of interest in the last two
decades \cite{Szek}-\cite{KonkHel}. Apart from the character of the
nonlinearities of the gravitational interaction, the interest was probably
caused by characteristic curvature singularities occuring as the result of the
collision of two waves. Much work has been done on the structure of the
singularities \cite{ChandraFer}-\cite{FerIban} in the case of collision of
either sourceless or various source waves, with the result that the final
singularity formation is, in fact, generic.

A relevant contribution concerning the singularity formation was made by
Hayward \cite{HayCQG}, who formulated the criterion of ``incoming''
regularity. In other words, he proposed to make a clear distinction whether
the singularity formation occurs for the collision of waves which are
initially regular or singular. Then the problem reads: Under what conditions
may the initially regular waves avoid the singularity formation after the
collision? For the case of the purely gravitational (sourceless) waves
Hayward himself found that the regular waves generically produce the curvature
singularities. However, there were also exceptional cases where the
singularities were avoided.

In D=2+1 the present authors found that after the collisions of regular
asymptotic scalar waves the singularity is always formed \cite{PrevPap}.
We considered the minimally coupled scalar field.
A similar conclusion was then obtained in D=3+1 by Hayward \cite{HayPrep}, who
has shown that if the pure gravitational radiation is absent in the initial
data then the collisions of the minimally coupled scalar waves always end up
in a singularity.

In this paper, we wish to study a source field of a different type and find
whether similar conclusions about the inevitability of the
singularity formation
can be reached. We shall work with the conformal scalar field with the field
equation (in the D-dimensional spacetime) \cite{Greek},
\cite{Klim}
\be \nabla_\a\nabla^\a\f-{{\rm D}-2\over{4({\rm D}-1)}}R\f=0,\label{FieldEq}
\ee
following from the action
$$S=\int\bigg[({1\over{2\k}}-{{\rm D}-2\over{8({\rm D}-1)}}\f^2)R-{1\over 2}
(\nabla\f)^2\bigg]\sqrt{-g}\,{\rm d^D}x.$$
Unlike the other massless field equations (i.e.~Maxwell, Dirac or Weyl), the
minimally coupled massless scalar equation is not conformally invariant. The
coupling according to (\ref{FieldEq}) cures this ``deficiency'' and, in any
case, it is a reasonable alternative for gravitational coupling of the
scalar field. The stress tensor for the conformal scalar field is quite
different from the ordinary one, and we may therefore test the singularity
formation problem in a different setting to previously.

{}From a technical point of view, it is not difficult to generate
solutions of the Einstein-conformal scalar equations from the minimally coupled
Einstein-scalar solutions via the generalized Bekenstein transformation
\cite{Greek}, \cite{Klim}. However, the structure of singularities requires
independent analysis, since the Bekenstein transformation multiplies the
original metric by a nontrivial conformal factor. This operation, in general,
{\em may\/} change the
asymptotic behaviour of the Riemann tensor components. Indeed, consider
for instance the dilaton gravity in D=2+1 with the action
$$ S=\int {\rm d}^3x\sqrt{-g}\,{\rm e}^\f R.$$
There is the following (black hole) solution of the corresponding field
equations \cite{KlimUnpubl}
\be
{\rm d}s^2=-{r-2m\over r}\,{\rm d}t^2+{r\over{r-2m}}\,{\rm d}r^2+r^2{\rm d}
\Th^2,\ \f=\log\abs{r\sin\Th}.\label{Schw}
\ee
On the other hand under the transformation
$$\wt g_{\a\b}={\rm e}^{2\f}g_{\a\b},\ \wt\f=\f,$$
the action $S$ changes into the action of the minimally coupled scalar field
$$ S=\int{\rm d}^3x\sqrt{-\wt g}\,(\wt R-2\wt\f_\g\wt\f^\g).$$
The metric corresponding to (\ref{Schw}) becomes
$${\rm d}s^2=r^2\sin^2\Th\left(-{r-2m\over r}\,{\rm d}t^2+{r\over{r-2m}}\,
{\rm d}r^2+r^2{\rm d}\Th^2\right).$$
It is not difficult to demonstrate that the singularity structures of both
metrics differ considerably from each other.

In sections 2 and 3
we study the singularity structure of the Einstein-conformal
scalar spacetimes in 2+1 and 3+1 dimensions, respectively. We arrive again to
the same conclusions as for the
minimally coupled scalar field, namely, for initially regular waves (without
sourceless part) the singularity formation  is inevitable.

\section{Conformal scalar waves in D=2+1}
\setcounter{equation}{0}

We start with a reminder of some properties of the spacetimes describing
the collision of line-fronted scalar waves. (It is known that the formation
of singularities occurs as the result of these processes \cite{PrevPap}.)
In such spacetimes, global coordinates can be introduced such that
the metric has the form
\be{\rm d}s^2=-{\rm e}^{-\wt K(u,v)}{\rm d}u\,{\rm d}v+{\rm e}^{-\wt N(u,v)}
{\rm d}x^2\label{Metric}\ee
with one space-like coordinate $x$ and a pair of null coordinates $u,\ v$.
The form of the metric functions $\wt N,\ \wt K$ follows from the Einstein
equations where on the r.h.s.~the stress tensor for the scalar field
$\wt\f$ stands:
\be\wt T_{\m\n}=\wt\f_\m\wt\f_\n-\half \wt g_{\m\n}\wt g^{\r\s}\wt\f_\r\wt
\f_\s.\label{Stress}\ee
Restricting ourselves to the case of the so-called asymptotic waves
spacetimes, for which all functions $\wt N,\ \wt K,\ \wt \f$ are smooth
and fulfil asymptotic conditions
\bea
(\wt K,\ \wt N,\ \wt\f)(u\to -\infty,\ v\to -\infty)&=&0,\nn\\
(\wt K,\ \wt N,\ \wt\f)(u,\ v\to -\infty)&=&(\wt K,\ \wt N,\ \wt\f)(u),\nn\\
(\wt K,\ \wt N,\ \wt\f)(u\to -\infty,\ v)&=&(\wt K,\ \wt N,\ \wt\f)(v),
\label{AC}
\eea
we can write down the general solution
$$\wt N=-2\ln\left(1-f(u)-g(v)\right),$$
\bea
\wt \f&=&k\ln(1-f-g)+p\cosh^{-1}\left[{1+f-g\over{1-f-g}}\right]
+q\cosh^{-1}\left[{1-f+g\over{1-f-g}}\right]\nn \\
&+&\int_0^\infty\!\!\left[A(\o)J_0\left(\o(1\!-\!f\!-\!g)\right)+B(\o)N_0
\left(\o(1\!-\!f\!-\!g)\right)\right]\sin\left(\o(f\!-\!g)\right){\rm d}
\o\nn\\
&+&\int_0^\infty\!\!\left[C(\o)J_0\left(\o(1\!-\!f\!-\!g)\right)+D(\o)N_0
\left(\o(1\!-\!f\!-\!g)\right)\right]\cos\left(\o(f\!-\!g)\right){\rm d}\o,
\nn\\
\label{Sol}
\eea
subject to
$$\int_0^\infty[C(\o)J_0(\o)+D(\o)N_0(\o)]\ {\rm d}\o=0,$$
where $f(u)$ and $g(v)$ are functions, $k,\ p$ and $q$ are real numbers,
$A(\o)$, $B(\o)$, $C(\o)$ and $D(\o)$ may be integrable functions or
distributions and $J_0$ and $N_0$ are zero-order Bessel and Neumann functions.

The function $\wt K$ can be obtained by integration from the relevant Einstein
equations, viz.
\bea
\wt N_{uu}-\half\wt N_u^2+\wt N_u\wt K_u&=&2\k\wt\f_u^2,\nn\\
\wt N_{vv}-\half\wt N_v^2+\wt N_v\wt K_v&=&2\k\wt\f_v^2,\nn\\
\wt K_{uv}&=&\k\wt\f_u\wt\f_v.\label{RelEin}
\eea
The asymptotic conditions (\ref{AC}), requiring certain asymptotic behaviour
of the functions $f(u)$ and $g(v)$, are met by the choice
\bea
f(u)&=&[-a(u-u_s)]^{1/(1-\k p^2)}\qquad{\rm for}\ \k p^2>1,\nn\\
f(u)&=&\exp[a(u-u_s)]\qquad\qquad\ {\rm for}\ \k p^2=1,\nn
\eea
and
\bea
g(v)&=&[-b(v-v_s)]^{1/(1-\k q^2)}\qquad{\rm for}\ \k q^2>1,\nn\\
g(v)&=&\exp[b(v-v_s)]\qquad\qquad\ {\rm for}\ \k q^2=1.\nn
\eea
Now, the plan of the investigation of the properties of these spacetimes was
the following \cite{PrevPap}: firstly we had to exclude the cases when the
asymptotic waves
were singular themselves, for the ``good physical situation'' should avoid
singularities in the initial data. Then the formation of final singularities
was questioned.
It turned out that in D=2+1, as the result of the collision of {\it regular\/}
asymptotic scalar waves, the final singularity always appeared \cite{PrevPap}.

Now, what is the situation when the source of colliding waves is not the
scalar field, but the conformal scalar field?
The solutions for the selfgravitating conformal scalar field can be easily
obtained from the selfgravitating minimal scalar spacetimes via a generalized
Bekenstein transformation \cite{Greek}, \cite{Klim}, linking the D-dimensional
scalar field
$\wt\f$ and metric $\wt g_{\a\b}$ to the D-dimensional conformal scalar field
$\f$ and metric $g_{\a\b}$ as follows
\bea
\f&=&\left(\k{\rm D-2\over{4(D-1)}}\right)^{-1/2} \tanh\left[
\left(\k{\rm D-2\over{4(D-1)}}\right)^{-1/2}\wt\f\right],\nn\\
g_{\a\b}&=&\left(\cosh\left[\left(\k{\rm D-2\over{4(D-1)}}\right)^{-1/2}
\wt\f\right]\right)^{4/({\rm D}-2)}\wt g_{\a\b}.\label{Bek}
\eea
In terms of metric functions $\wt N$, $\wt K$ and analogously defined $N$
and $K$ we have (for D=2+1) the following transformation rules
\bea
\f&=&\sqrt{8\over\k}\tanh\left[\sqrt{\k\over 8}\wt\f\right],\nn\\
N&=&\wt N-4\ln\cosh\left[\sqrt{\k\over 8}\wt\f\right],\nn\\
K&=&\wt K-4\ln\cosh\left[\sqrt{\k\over 8}\wt\f\right].\label{Trans}
\eea
We see that (\ref{AC}) imply fulfillment of the same asymptotic conditions
for the new metric functions and the new (conformal scalar) field. It means
that the metric $g_{\a\b}$ also describes the colliding asymptotic waves
spacetimes and we can study the conditions for the regularity of initial
data as well as the consequent creation of singularities after
collision\footnote{Actually, there is one more branch of the generalized
Bekenstein transformation (see \cite{Klim}) which, however, does
not preserve the asymptotic conditions, so we shall not deal with it
anymore.}.
What is the criterion for the regularity of the initial data?

Consider the (asymptotic) regions where only one of the initial waves is
present
while the influence of the other wave vanishes, i.e.~the regions ($u\to-
\infty$, $v\ne-\infty$) or ($v\to -\infty$, $u\ne-\infty$). Now the metric
is that of the single wave there (i.e.~$u$- or $v$-independent) and we can
introduce the amplitude ${}^uh(u)$
$${}^uh(u)=\half{\rm e}^{2K}(\half N_u^2-N_{uu}-N_uK_u),$$
resp.~${}^vh(v)$
$${}^vh(v)=\half{\rm e}^{2K}(\half N_v^2-N_{vv}-N_vK_v).$$

These are the amplitudes of the single line-wave in the so-called Brinkmann
coordinates. It was shown in \cite{PrevPap} that the boundedness of these
amplitudes is the necessary and sufficient condition for the absence of the
curvature singularities in the asymptotic (incoming) region.

Using the transformation rules (\ref{Trans}) and the Einstein equations for
the ordinary scalar field (\ref{RelEin}) we can write
$${}^uh(u)=\half{\rm CH}^{-8}{\rm e}^{2\wt K}\left(\k\wt\f_u^2(-2-
{\rm TH}^2+{1\over{2{\rm CH}^2}})+\sqrt{2\k}\,{\rm TH}\,(\wt\f_{uu}+
\wt\f_u\wt K_u)\right),$$
where CH=$\cosh(\wt\f\sqrt{\k/8})$, TH=$\tanh(\wt\f\sqrt{\k/8})$ and the
other amplitude ${}^vh(v)$ has the analogous form. Now, we need the asymptotic
behaviour of the Bessel and Neumann functions near zero
\bea
J_0(w)&\sim& 1-{w^2\over 4}+\cdots,\nn\\
J'_0(w)&\sim& -{w\over 2}+\cdots,\nn\\
N_0(w)&\sim& (1-{w^2\over 4})\ln w+\cdots,\nn\\
N'_0(w)&\sim& {1\over w}-{w\over 2}\ln w+\cdots\label{Expan}
\eea
{}From the expressions (\ref{Sol}) taken in $g=0$, $f\sim 1$ it follows that
$$\wt\f_u\sim f_u\left[{c\over 1-f}+d\ln (1-f)+e+h(1-f)\ln (1-f)+
\cdots\right],$$
and then from (\ref{RelEin}) we can obtain
$$\wt K_u=f_u\left[{\k c^2\over{1-f}}+2\k cd\ln(1-f)+\cdots\right],$$
and hence
$$\wt K=-\k c^2\ln(1-f)+{\textstyle\rm bounded}.$$
Here
\bea
c&=&p-k-\int_0^\infty\!\![B(\o)\sin(\o)+D(\o)\cos(\o)]{\rm d}\o,\nn\\
d&=&\int_0^\infty\!\![\o B(\o)\cos(\o)-\o D(\o)\sin(\o)]{\rm d}\o,\nn\\
e&=&{p\over 2}+\int_0^\infty\!\!\o[A(\o)+B(\o)\ln\o]\cos(\o){\rm d}\o
\nn\\
& &-\int_0^\infty\!\!\o[C(\o)+D(\o)\ln\o]\sin(\o){\rm d}\o,\nn\\
h&=&\half\int_0^\infty\!\!\o^2[B(\o)\sin(\o)+D(\o)\cos(\o)]{\rm d}\o.
\label{Coef}
\eea
Thus if $c\ne 0$ the leading term in ${}^uh(u)$ is
$$2^7 f_u^2(1-f)^{8\abs c\sqrt{\k/8}-2\k c^2-2}\abs c\left[-3\k \abs c
+\sqrt{2\k}(1+\k c^2)\right].$$
It vanishes for $\abs c=1/\sqrt{2\k}$ or $\abs c=2/\sqrt{2\k}$,
but then we cannot suppress the subleading term  proportional to
 $(1-f)^{-1/2}$ or $(1-f)^{-1}$ respectively. Hence, we arrived
at the first {\it necessary\/} condition for the incoming regularity,
viz.~$c=0$. In this case the asymptotic behaviour of ${}^uh(u)$ is
$${}^uh(u)\sim{\textstyle\rm bounded}\times\left[-\half\k
(3d^2\ln^2(1-f)+(6de-d^2)\ln(1-f))+\cdots\right],$$
from which the second necessary condition for the incoming regularity can be
extracted: $d=0$. It is easy to see that the conditions $c=d=0$ are also
sufficient.

After a similar analysis for ${}^vh(v)$ we conclude that for the incoming
regularity it is necessary and sufficient to fulfil
\be c=d=c^\prime=d^\prime=0,\label{IncReg}\ee
where $c$ and $d$ are as above and
\bea
c^\prime&=&q-k+\int_0^\infty\!\![B(\o)\sin(\o)-D(\o)\cos(\o)]{\rm d}\o,\nn\\
d^\prime&=&-\int_0^\infty\!\![\o B(\o)\cos(\o)+\o D(\o)\sin(\o)]{\rm d}\o.\nn
\eea

Now, we wish to study the region of interaction, i.e.~the region where the
metric is both $u$- and $v$-dependent. There are special points (forming
the so-called caustic) in which the metric functions are singular, namely
the points where
$$1-f-g=0.$$
We have to elucidate what kind of singularity this is, or -- in
other words -- how the curvature behaves in its vicinity.
In particular, we find the conditions under which the scalar curvature $R$
is unbounded while approaching the caustic\footnote{This case is often
referred to as the $C^0$ scalar curvature singularity (see
\cite{Ellis+Schmidt}).}.
In terms of metric functions the scalar curvature $R$ reads
$$R=-4{\rm e}^K K_{uv},$$
which we can rewrite using (\ref{Trans}) and (\ref{RelEin})
$$R=-4\,{\rm e}^{\wt K}{\rm CH}^{-4}[\k\wt\f_u\wt\f_v-\sqrt{2\k}\,\wt\f_{uv}
{\rm TH}-\half\k\wt\f_u\wt\f_v{\rm CH}^{-2}].$$
In the neighbourhood of the caustic it is convenient to introduce other
functions $t$ and $z$ instead of $f$ and $g$, given by
\be t=1-f-g,\qquad z=f-g.\label{Functions}\ee
Then the caustic is formed by the points $t=0, \abs z\ne 1$, the second
condition guaranteeing that we do not deal with the asymptotic caustic studied
previously. Near the caustic
the following behaviour of $\wt K$ and $\wt\f$ can be derived from (\ref{Sol})
and (\ref{Expan})
\bea\wt\f&\sim&E(z)\ln t+\cdots,\nn\\
\wt K&\sim&-\k\,E^2\!(z)\ln t+\cdots,\nn \eea
where
\be E(z)=k-p-q+\int_0^\infty\!\![B(\o)\sin(\o z)+D(\o)\cos(\o z)]
{\rm d}\o.\label{E}\ee
So unless $E(z)\equiv 0$ the leading term in $R$ is
$$ f_ug_v\,{\textstyle\rm const}\ t^{-\k E^2\!(z)+4\abs{E(z)}\sqrt{\k/8}-2}
[\k E^2\!(z)-\sqrt{2\k}\left(\sign E(z)\right)E(z)],$$
which vanishes if $\abs{E(z)}=\sqrt{2/\k}$. But then the subleading term
$$-4\,f_ug_v\,{\textstyle\rm const}\ t^{-1}$$
is present in $R$, so $R$ is unbounded. Hence the only case when $R$
{\em may} be bounded near the caustic is if $E(z)\equiv 0$. (This, in turn,
can only be true for $B(\o)=D(\o)=0$ because of the nontrivial dependence
of $E(z)$ on $z$.)

However, the condition $E=0$ is incompatible with the conditions
(\ref{IncReg}) for the asymptotic (incoming) regularity. We can conclude that
after collision of regular asymptotic conformal scalar waves the scalar
curvature singularity at the caustic is always produced. Hence, the conclusion
is the same as in the case of minimal scalar waves.

\section{Conformal scalar waves in D=3+1}
\setcounter{equation}{0}

In this section, we shall proceed to the collisions of asymptotic
conformal scalar
waves in D=3+1. In order to obtain a solution which describes
such a process, we again apply  the Bekenstein transformation (\ref{Bek}).
The metric with the minimally coupled scalar field $\wt\f$ as the
source\footnote{Unfortunately,
the exact solutions with scalar sources are known only for the case of
so-called
collinear waves, it means $\wt g_{xy}=0$ in terms of the metric given below.}
has the form (see \cite{HayPrep})
$${\rm d}s^2=-2{\rm e}^{-\wt M}{\rm d}u{\rm d}v+{\rm e}^{-\wt P+\wt Q}{\rm d}
x^2+{\rm e}^{-\wt P-\wt Q}{\rm d}y^2.$$
The metric functions and $\wt\f$, fulfilling the asymptotic conditions
analogous to (\ref{AC}), can be expressed in the form
$$\wt P=-\ln\left(1-f(u)-g(v)\right),$$
\bea
\wt Q&=&k\ln(1-f-g)+p\cosh^{-1}\left[{1+f-g\over{1-f-g}}\right]
+q\cosh^{-1}\left[{1-f+g\over{1-f-g}}\right]\nn \\
&+&\int_0^\infty\!\!\left[A(\o)J_0\left(\o(1\!-\!f\!-\!g)\right)+B(\o)N_0
\left(\o(1\!-\!f\!-\!g)\right)\right]\sin\left(\o(f\!-\!g)\right){\rm d}
\o\nn\\
&+&\int_0^\infty\!\!\left[C(\o)J_0\left(\o(1\!-\!f\!-\!g)\right)+D(\o)N_0
\left(\o(1\!-\!f\!-\!g)\right)\right]\cos\left(\o(f\!-\!g)\right)
{\rm d}\o,\nn
\eea
\bea
\wt \f&=&\l\ln(1-f-g)+\p\cosh^{-1}\left[{1+f-g\over{1-f-g}}\right]
+\x\cosh^{-1}\left[{1-f+g\over{1-f-g}}\right]\nn\\
&+&\int_0^\infty\!\!\left[{\cal A}(\o)J_0\left(\o(1\!-\!f\!-\!g)\right)+{\cal
B}(\o)N_0\left(\o(1\!-\!f\!-\!g)\right)\right]\sin\left(\o(f\!-\!g)\right)
{\rm d}\o\nn\\
&+&\int_0^\infty\!\!\left[{\cal C}(\o)J_0\left(\o(1\!-\!f\!-\!g)\right)+{\cal
D}(\o)N_0\left(\o(1\!-\!f\!-\!g)\right)\right]\cos\left(\o(f\!-\!g)\right)
{\rm d}\o,\nn
\eea
subject to the constraints
\bea
\int_0^\infty\!\![C(\o)J_0(\o)+D(\o)N_0(\o)]{\rm d}\o=0,\nn\\
\int_0^\infty\!\![{\cal C}(\o)J_0(\o)+{\cal D}(\o)N_0(\o)]{\rm d}\o=0.\nn
\eea
All symbols have analogous meaning as in the D=2+1 case. The last --
unexpressed -- metric function $\wt M$ is given by direct integration of
the relevant Einstein equations
\bea
2\wt P_{uu}-\wt P_u^2+2\wt P_u\wt M_u&=&\wt Q_u^2+2\k\wt\f_u^2,\nn\\
2\wt P_{vv}-\wt P_v^2+2\wt P_v\wt M_v&=&\wt Q_v^2+2\k\wt\f_v^2,\nn\\
2\wt M_{uv}&=&\wt Q_u\wt Q_v-\wt P_u\wt P_v+2\k\wt\f_u\wt\f_v,\label{RelEin4}
\eea
(in our conventions the stress tensor is double of that considered
by Hayward \cite{HayPrep}).
In what follows we shall restrict ourselves to the special type of functions
$f(u)$ (and $g(v)$) \cite{HayPrep}
\bea
f(u)&=&[-a(u-u_s)]^{2/(2-p^2-2\k\p^2)}\qquad{\rm for}\ p^2+2\k\p^2>2,\nn\\
f(u)&=&\exp[a(u-u_s)]\qquad\qquad\qquad\ {\rm for}\ p^2+2\k\p^2=2,\nn
\eea
ensuring the proper asymptotic behaviour of the metric functions and of
the field in past timelike infinity.
Then the Bekenstein transformation (\ref{Bek}) yields the new metric
functions $M$, $P$ and $Q$ and new (conformal scalar) field $\f$:
\bea
\f&=&\sqrt{6\over\k}\tanh\left[\sqrt{\k\over 6}\wt\f\right],\nn\\
P&=&\wt P-2\ln\cosh\left[\sqrt{\k\over 6}\wt\f\right],\nn\\
M&=&\wt M-2\ln\cosh\left[\sqrt{\k\over 6}\wt\f\right],\nn\\
Q&=&\wt Q.\label{Trans4}
\eea
We turn to the study of the singularity structure of the new spacetimes
(\ref{Trans4}).

Near the asymptotic caustic, e.g.~($v=-\infty$, $u=0$), where the metric is
$v$-independent, it is convenient to take a new null coordinate $u^\prime$
instead of $u$ such that
$${\rm d}u^\prime={\rm e}^{-M}{\rm d}u.$$
Then the following vierbein is orthonormal and parallelly propagated along
the incomplete geodesics respecting $x$- and $y$-symmetry and hitting the
asymptotic caustic:
\bea
e^\a_{(0)}&=&({1\over{\sqrt 2}},{1\over{\sqrt 2}},0,0),\nn\\
e^\a_{(1)}&=&({1\over{\sqrt 2}},-{1\over{\sqrt 2}},0,0),\nn\\
e^\a_{(2)}&=&(0,0,{\rm e}^{\half(P-Q)(u^\prime)},0),\nn\\
e^\a_{(3)}&=&(0,0,0,{\rm e}^{\half(P+Q)(u^\prime)}).\nn
\eea
It is straightforward to compute the only two (mutually independent)
nonzero vierbein components of the Riemann curvature tensor as
\bea
R_{2020}&=&-{\textstyle 1\over 8}[2Q_{u^\prime u^\prime}-
2P_{u^\prime u^\prime}+(Q_{u^\prime}-P_{u^\prime})^2],\nn\\
R_{3030}&=&{\textstyle 1\over 8}[2Q_{u^\prime u^\prime}+
2P_{u^\prime u^\prime}-(Q_{u^\prime}+P_{u^\prime})^2].\nn
\eea
Going back to the original coordinates we have
\bea
R_{2020}&=&-{\textstyle 1\over 8}{\rm e}^{2M}[2Q_{uu}+2Q_uM_u-
2P_{uu}-2P_uM_u+(Q_u-P_u)^2],\nn\\
R_{3030}&=&{\textstyle 1\over 8}{\rm e}^{2M}[2Q_{uu}+2Q_uM_u+
2P_{uu}+2P_uM_u-(Q_u+P_u)^2].\nn
\eea
The incoming regularity now requires the boundedness of both components.
Using (\ref{Trans4}), we can write
\bea
{}^uR^-&\equiv&R_{2020}-R_{3030}=-\half{\rm Ch}^{-4}\,{\rm e}^{2\wt M}
[\wt Q_{uu}+\wt Q_u\wt M_u
-\wt Q_u\wt P_u],\nn\\
{}^uR^+&\equiv&R_{2020}+R_{3030}={\textstyle 1\over4}{\rm Ch}^{-4}\,
{\rm e}^{2\wt M}
[2\wt P_{uu}-4\sqrt{\k\over 6}\wt\f_{uu}{\rm Th}-{\textstyle 2\over 3}\k
\wt\f_u^2{\rm Ch}\nn\\
& &+(\wt P_u-2\sqrt{\k\over 6}\wt\f_u{\rm Th})(2\wt M_u-\wt P_u-2\sqrt{\k
\over 6}\wt\f_u{\rm Th})-\wt Q_u^2],\nn
\eea
with Ch and Th standing instead of $\cosh(\wt\f\sqrt{\k/6})$ and $\tanh(\wt\f
\sqrt{\k/6})$, respectively.

We have to identify the behaviour of ${}^u R^\pm$ for $f\sim 1$. Taking
again into account the asymptotic behaviour of the Bessel and Neumann
functions (\ref{Expan}), we obtain (for $f\sim 1$):
\bea
\wt Q_u&\sim&f_u\left[{c\over 1-f}+d\ln (1-f)+e+h(1-f)\ln (1-f)+
\cdots\right],\nn\\
\wt\f_u&\sim&f_u\left[{c^*\over 1-f}+d^*\ln (1-f)+e^*+h^*(1-f)\ln (1-f)+
\cdots\right],\nn
\eea
where {\it c, d, e, h} have the same form as in (\ref{Coef}) and $c^*$, $d^*$,
$e^*$, $h^*$ are given by
\bea
c^*&=&\p-\l-\int_0^\infty\!\![{\cal B}(\o)\sin(\o)+{\cal D}(\o)\cos(\o)]
{\rm d}\o,\nn\\
d^*&=&\int_0^\infty\!\![\o {\cal B}(\o)\cos(\o)-\o {\cal D}(\o)\sin(\o)]
{\rm d}\o,\nn\\
e^*&=&{\p\over 2}+\int_0^\infty\!\!\o[{\cal A}(\o)+{\cal B}(\o)\ln\o]
\cos(\o){\rm d}\o\nn\\
& &-\int_0^\infty\!\!\o[{\cal C}(\o)+{\cal D}(\o)\ln\o]\sin(\o){\rm d}\o,\nn\\
h^*&=&\half\int_0^\infty\!\!\o^2[{\cal B}(\o)\sin(\o)+{\cal D}(\o)\cos(\o)]
{\rm d}\o.
\eea
{}From (\ref{RelEin4}) we have
$$\wt M_u=f_u\left[{c^2+2\k{c^*}^2-1\over 2}{1\over {1-f}}+(cd+2\k c^*d^*)
\ln(1-f)+\cdots\right],$$
hence
$$\wt M=-{c^2+2\k{c^*}^2-1\over 2}\ln(1-f)+{\textstyle\rm bounded}.$$
Therefore, the leading singular terms in ${}^uR^-$ and ${}^uR^+$ read
\bea
-8f_u^2\,{c^2+2\k{c^*}^2-1\over 2}c\,(1-f)^{4\sqrt{\k/6}\abs{c^*}-
(c^2+2\k{c^*}^2-1)-2},\nn\\
4f_u^2\,\abs{c^*}\left[{8\k\over3}\abs{c^*}-2\sqrt{\k\over 6}
(c^2+2\k{c^*}^2+1)\right]\,(1-f)^{4\sqrt{\k/6}\abs{c^*}-
(c^2+2\k{c^*}^2-1)-2}.\nn
\eea
Both coefficients of proportionality vanish if
\bea
c=0,\ \abs{c^*}=(2\pm 1)/\sqrt{6\k},\nn\\
\abs c=\half,\ \abs{c^*}=\sqrt{3\over{8\k}},\nn\\
c=c^*=0,\nn\\
c^*=0,\ \abs c=1.\nn
\eea
In the first two cases there are subleading singular terms, which cannot
be eliminated, but in the remaining cases we can exclude them by fitting some
other coefficients in the expansions of $\wt\f$ and $\wt Q$.
Hence, the initial data are free of curvature singularities if $c=c^*=d=0$ or
if $c^2=1$, $c^*=d=d^*=h=0$. The analogous conditions have to be satisfied
at the other asymptotic caustic.

Now, we have to study the components of the curvature tensor at the caustic
$1-f-g=0$. It is sufficient \cite{HayPrep} to consider the scalar curvature
$R$ given by
$$R=-{\rm e}^M(P_uP_v+2M_{uv}-Q_uQ_v),$$
and the component $\Psi_2$ of the Weyl spinor in the null spin-frame
\cite{HayPrep}
$$\Psi_2={\textstyle 1\over 3}{\rm e}^M(Q_uQ_v-P_uP_v+M_{uv}).$$
If one of them is unbounded, then there is a (final) curvature singularity
at the caustic. We take the suitable combinations of $\Psi_2$ and $R$:
\bea
V_1&=&{\rm e}^M M_{uv},\nn\\
V_2&=&{\rm e}^M (Q_uQ_v-P_uP_v).\nn
\eea
Using (\ref{Trans4}) we have
\bea
V_1&=&{\rm Ch}^{-2}{\rm e}^{\wt M}\left[\wt M_{uv}-2\sqrt{\k\over 6}\wt\f_{uv}
{\rm Th}-{\textstyle 1\over 3}\k\wt\f_u\wt\f_v{\rm Ch}^{-2}\right],\nn\\
V_2&=&{\rm Ch}^{-2}{\rm e}^{\wt M}\left[\wt Q_u\wt Q_v-\bigl(\wt P_u-2
\sqrt{\k\over 6}\wt\f_u{\rm Th}\bigr)\bigl(\wt P_v-2\sqrt{\k\over 6}\wt\f_v
{\rm Th}\bigr)\right].\nn
\eea
We again introduce the functions $t$ and $z$ as in (\ref{Functions}).
The asymptotic behaviour of $\wt\f$ and $\wt Q$ near $t=0$ is
\bea
\wt\f&\sim&{\cal E}(z)\ln t+{\cal F}(z)t^2,\nn\\
\wt Q&\sim&E(z)\ln t+F(z)t^2,\nn
\eea
where
$${\cal E}(z)=\l-\p-\x+\int_o^\infty\!\![{\cal B}(\o)\sin(\o z)+{\cal D}(\o)
\cos(\o z)]{\rm d}\o,$$
$E(z)$ is given as (\ref{E}) and the forms of $F(z)$ and ${\cal F}(z)$ are
not important.
Then the leading singular terms of $V_1$, $V_2$ are proportional to
\bea
\left[E^2\!(z)+2\k{\cal E}^2\!(z)-1-4\sqrt{\k\over 6}\abs{{\cal E}(z)}
\right]t^{2\sqrt{\k/6}\abs{{\cal E}(z)}+\half(1-E^2\!(z)-2\k
{\cal E}^2\!(z))-2},\nn\\
\left[E^2\!(z)-{\textstyle 2\over 3}\k{\cal E}^2\!(z)-1+4\sqrt{\k\over 6}
\abs{{\cal E}(z)}\right]t^{2\sqrt{\k/6}\abs{{\cal E}(z)}+\half(1-E^2\!(z)
-2\k{\cal E}^2\!(z))-2},\nn
\eea
respectively. Both coefficients of proportionality vanish if
\bea
{\cal E}=0,\ \abs E=1,\nn\\
E=0,\ \abs{{\cal E}}=\sqrt{3\over{2\k}}.\nn
\eea
In the second case, $V_1$ is unbounded due to the subleading term. Hence the
only way to keep both $V_1$ and $V_2$ bounded is to set ${\cal E}(z)\equiv 0$
and $\abs{E(z)}\equiv 1$.

Therefore both the criterion for the incoming regularity and the necessary
condition for the avoiding of the final singularity are in the case of
conformal scalar waves the same as in the case of minimal scalar
waves
\cite{HayPrep}. Hence the conclusions have to be the same, too. In particular,
the formation of final singularities in the collision of regular asymptotical
conformal scalar waves is generic. Moreover, if the pure gravitational
radiation is absent in the initial data, i.e.
\bea
(-Q_{uu}+P_uQ_u-M_uQ_u)=0,\ v\to -\infty,\nn\\
(-Q_{vv}+P_vQ_v-M_vQ_v)=0,\ u\to -\infty,\nn
\eea
the final singularities are even inevitable.

\section{Concluding remarks}

In 2+1 dimensions, where there is no pure gravitational radiation, the only
gravitational waves are those accompanying light-like matter sources.
For the massless minimally coupled scalar field \cite{PrevPap} previously
and for the conformal scalar field now, we have shown that collisions of
regular
waves necessarily form the final curvature singularities. Since the
electrovacua in D=2+1 are (up to some global obstructions of the
cohomological origin) equivalent to massless minimally coupled scalar vacua
\cite{KlimUnpub}, it is reasonable to conjecture that collisions of
regular general light-like matter waves always end up in a curvature
singularity.
Of course, systems with more exotic stress tensors have to be
studied in order to prove this conjecture. From the point of view of quantum
theory it may be worth remarking that the classical phase space, which is to
be quantized, does not possess too complicated a structure from the geometrical
point of view. Indeed, the spacetimes for all regular initial data have the
same global structure.

In 3+1 dimensions the situation is slightly more complicated due to the
presence of the pure (sourceless) gravitational radiation. Indeed, within the
framework of pure gravity there are regular initial data the evolution of
which does not lead to a curvature singularity at the caustic. However, if we
exclude the sourceless waves in the initial data,
we have shown that the collision of the regular pure
conformal scalar waves inevitably leads to the formation of a curvature
singularity. A similar conclusion was made for the case of the massless
minimally coupled scalar field \cite{HayPrep}.

\vskip .7cm
{\large\bf\noindent Acknowledgment}
\vskip .5cm

\noindent We are grateful to Sean Hayward for enlightening discussions.

\end{document}